\documentclass[12pt,notoc]{JHEP3}

\pdfoutput=1
\usepackage{bm}
\usepackage{epsfig}
\usepackage{citesort}
\usepackage{graphicx}

\def\be{\begin{equation}}
  \def\ee{\end{equation}}
\def\bea{\begin{eqnarray}}
  \def\eea{\end{eqnarray}}

\preprint{LAPTH-1350/09}

\title{Features in the primordial power spectrum? A frequentist
analysis} 

\author{Jan Hamann\\
	Department of Physics and Astronomy, University of Aarhus\\ 
	8000 \AA rhus C, Denmark\\
	\emph{and} \\
	LAPTh, Universit\'e de Savoie, CNRS\\ 
	BP 110, 74941 Annecy-le-Vieux Cedex, France\\
	E-mail: \email{hamann@phys.au.dk}}

\author{Arman Shafieloo\\
	Department of Physics, University of Oxford\\ 
	1 Keble Road, Oxford, OX1 3NP, UK\\
	E-mail: \email{a.shafieloo1@physics.ox.ac.uk}}

\author{Tarun Souradeep\\
        IUCAA\\
        Post Bag 4, Ganeshkhind, Pune 411 007, India\\
        E-mail: \email{tarun@iucaa.ernet.in}}

\keywords{CMBR theory, cosmological parameters from CMBR, initial
conditions and eternal universe} 
     
\abstract{Features in the primordial power spectrum have been
  suggested as an explanation for glitches in the angular power
  spectrum of temperature anisotropies measured by the WMAP satellite.
  However, these glitches might just as well be artifacts of noise or
  cosmic variance.  Using the effective $\Delta \chi^2$ between the
  best-fit power-law spectrum and a deconvolved primordial spectrum as
  a measure of ``featureness'' of the data, we perform a full
  Monte-Carlo analysis to address the question of how significant the
  recovered features are.  We find that in 26\% of the simulated data
  sets the reconstructed spectrum yields a greater improvement in the
  likelihood than for the actually observed data.  While features
  cannot be categorically ruled out by this analysis, and the
  possibility remains that simple theoretical models which predict
  some of the observed features might stand up to rigorous statistical
  testing, our results suggest that WMAP data are consistent with
  the assumption of a featureless power-law primordial spectrum.}

\begin{document}

%%%%%%%%%%%%%%%%%%%%%%%%%%%%%%%%%%%%%%%%%%%%%%%%%%%%%%%%%%%%%%%%%%%%%%
\section{Introduction}                        
\label{sec:introduction}
%%%%%%%%%%%%%%%%%%%%%%%%%%%%%%%%%%%%%%%%%%%%%%%%%%%%%%%%%%%%%%%%%%%%%%

The advent of a multitude of cosmological precision data in the past
decade has led to the emergence of the so-called concordance, or
``vanilla'' model of cosmology.  A key part of the vanilla model is
the assumption that the spectrum of primordial curvature perturbations
is smooth and featureless, and can be described by a simple power-law
-- consistent with their origin from an earlier period of slow-roll
inflation~\cite{Starobinsky:1982ee,Guth:1982ec,Bardeen:1983qw,Mukhanov:1990me}.

Nonetheless, the inflationary mechanism also allows for more complex
shapes of the spectrum, caused for instance by non-standard initial
conditions~\cite{Martin:1999fa,Martin:2000xs,Danielsson:2002kx,Contaldi:2003zv},
or unusual dynamics of the inflaton field due to, e.g., a phase
transition~\cite{Adams:1997de,Joy:2007na},
non-smoothness of the inflaton potential~\cite{Starobinsky:1992ts,Adams:2001vc,Jain:2008dw}
or particle production~\cite{Chung:1999ve,Elgaroy:2003hp} during
inflation (see also
\cite{Kofman:1986wm,KofPog1988,Salopek:1988qh,Polarski:1992dq,Kaloper:2003nv,Cline:2006db,Jain:2009pm}).
In any case, any detection of features in the spectrum, i.e., a
deviation from the standard power-law behaviour, would yield
invaluable clues on the physics of the early Universe.
Presently, the most powerful single source of information about the
primordial state of perturbations are the observations of the
temperature anisotropies of the Cosmic Microwave Background (CMB) by
the WMAP satellite~\cite{Nolta:2008ih,Dunkley:2008ie,Komatsu:2008hk}.
Since the first data release, a number of efforts have been undertaken
to ascertain the compatibility of the data with the power-law
paradigm.

In a top-down approach, specific non-smooth models of the primordial
spectrum or the inflaton potential have been fit to the data
\cite{Barriga:2000nk,Peiris:2003ff,Martin:2003sg,Martin:2004iv,Martin:2004yi,Hunt:2004vt,Bridges:2005br,Spergel:2006hy,Covi:2006ci,Hamann:2007pa,Hunt:2007dn,Joy:2008qd,Barnaby:2009dd,Ichiki:2009xs}. 
Alternatively, the bottom-up approach of trying to reconstruct the
shape of the primordial spectrum from these data has been employed,
involving for example binning of the primordial spectrum
\cite{Bridle:2003sa,Hannestad:2003zs,Mukherjee:2003ag}, principal
component analysis \cite{Leach:2005av}, or a direct reconstruction
via deconvolution
methods~\cite{Kogo:2003yb,Shafieloo:2003gf,TocchiniValentini:2004ht,Shafieloo:2006hs,Shafieloo:2007tk,Nagata:2008tk,Nagata:2008zj,Ichiki:2009zz,Nicholson:2009pi,Nicholson:2009zj}.

Generally, these results have shown that by introducing suitable
features in the primordial spectrum, the fit to the data can be
improved by $\Delta \chi^2_{\rm eff} \sim \mathcal{O}(10)$ over the
power-law fit.  Taken by itself, however, $\Delta \chi^2_{\rm eff}$
does not really help answer the crucial question whether the better
fit indicates a real feature in the primordial spectrum or whether it
just stems from over-fitting the scatter in the data from noise and
cosmic variance.  Indeed, attempts at interpreting results of
parameter inference in terms of 'ruling out the power-law model in
favour of a feature model' are likely to be prone to grossly
overestimating the 'evidence' for features.
A simplistic goodness-of-fit $\chi^2/{\rm dof}$ analysis would not be
very enlightening either, due to the large number of degrees of
freedom involved.  One might thus be tempted to turn to the formalism
of Bayesian model selection \cite{Trotta:2005ar,Liddle:2009xe}.
Though aside of the potential technical problems of evaluating Bayes
factors for feature models with a large number of free parameters, the
interpretation of results from a Bayesian model selection analysis may
be problematic \cite{Efstathiou:2008ed}, owing to the absence of a
unique well-motivated choice of priors for these mostly empirical
parameterisations.

We believe that for this particular problem the toolbox of frequentist
statistics provides a useful implement, and propose to perform an
analysis based on the technique of hypothesis testing.  Given the null
hypothesis of an underlying power-law spectrum, this involves the
evaluation of a suitably chosen statistic on a large number of Monte
Carlo simulated mock CMB temperature anisotropy data sets.  From the
resulting frequency distribution of this statistic one can derive a
$p$-value, i.e., the probability that, given the null hypothesis, the
value of the statistic is larger than the one observed.  We note that
in defining the statistic one has to beware of a posteriori
interpretations of the data; a particular feature observed in the real
data may be very unlikely (and lead to a low $p$-value; see e.g.,
\cite{Spergel:2003cb,Lewis:2003qm}), but the probability of observing
some feature may be quite large.
%{\bf One of the 
%important characteristics of this approach is that we can confront the theoretical 
%model (here it is power-law form of the primordial spectrum) directly to the data for a consistency check
%without involving any alternative model for comparison. } 

In Section~\ref{sec:method} (and in the Appendix) we describe in
detail the methods used in our analysis, the results of which are
presented in Section~\ref{sec:results}.  We discuss and interpret our
findings in Section~\ref{sec:conclusion}.

%%%%%%%%%%%%%%%%%%%%%%%%%%%%%%%%%%%%%%%%%%%%%%%%%%%%%%%%%%%%%%%%%%%%%%
\section{Method}                        
\label{sec:method}
%%%%%%%%%%%%%%%%%%%%%%%%%%%%%%%%%%%%%%%%%%%%%%%%%%%%%%%%%%%%%%%%%%%%%%

\subsection{Null hypothesis}

We assume as our null hypothesis that the temperature angular power
spectrum inferred from WMAP can be explained by the present
cosmological standard (``vanilla'') model. 
In particular, the underlying primordial spectrum of curvature
perturbations is taken to have a smooth power-law form
\begin{equation}
	\mathcal{P_R}(k) = A_{\rm S} \, \left(k/k_*\right)^{n_{\rm S}
	- 1},
\end{equation}
determined by the two parameters $n_{\rm S}$, the scalar spectral
index, and $A_{\rm S}$, the amplitude of fluctuations at the pivot
scale $k_* = 0.05 {\rm ~Mpc}^{-1}$.  The remaining four free
parameters of the vanilla model are the baryon and cold dark matter
densities $\omega_{\rm b}$ and $\omega_{\rm c}$, the ratio of sound
horizon to angular diameter distance at decoupling $\theta$, and the
reionisation optical depth $\tau$.

The fiducial spectrum $\mathcal{C}_\ell^{\rm fid}$ used to generate
random realisations of WMAP5 data should be chosen as the maximum
likelihood spectrum.  However, since temperature data alone are not
very sensitive to $\tau$, using temperature data alone leads to a
value of $\tau$ which is at odds with results of a combined
temperature/polarisation analysis.  We therefore fix $\tau$ to the
best-fit result for the full WMAP5 likelihood function, and then
determine the values of the other five parameters by fitting the thus
reduced five-parameter vanilla model to the temperature data only
(using the likelihood function described in Appendix~\ref{app:like}).

The resulting fiducial spectrum is defined by the following parameter
values: $\omega_{\rm b} = 0.0224 $, $\omega_{\rm c} = 0.109$, $\theta
= 1.04$, $\tau = 0.089$, $A_{\rm S} = 2.147 \cdot 10^{-9}$ and $n_{\rm
S} = 0.963$.

\subsubsection{Alternative hypothesis}

The null hypothesis is to be tested against an alternative
hypothesis. Our alternative hypothesis is that the primordial spectrum
is not given by a power-law, but possesses features of some sort.  It
now remains to define a suitable test statistic that can be used to
assess the validity of the null hypothesis.

\subsection{Statistic}

By relaxing the assumption of a precise power-law form of the
primordial spectrum one can generally achieve a better fit to the
data.  A commonly used measure of this improvement is the
effective delta-chi-squared
\begin{equation}
\Delta \chi^2_{\rm eff} = - 2 \ln \mathcal{L}_{\rm max}^{\rm V} + 2
\ln \mathcal{L}_{\rm max}^{X}, 
\end{equation}
where $\mathcal{L}_{\rm max}^{\rm V}$ is the maximum likelihood of a
fit to the vanilla model, and $\mathcal{L}_{\rm max}^{\rm X}$ the
maximum likelihood of a fit to a model with an alternative shape of
the spectrum.

In the present work we wish to allow a very general form of the
primordial spectrum, so we focus on a reconstruction of the primordial
spectrum from a binned version of the temperature angular power
spectrum with the aid of a modified Richardson-Lucy (RL) deconvolution
algorithm as performed in
Refs.~\cite{Shafieloo:2003gf,Shafieloo:2007tk} (see
Appendix~\ref{app:rl}).  Given a transfer function $T_\ell (k)$ and an
observed or simulated angular power spectrum $\mathcal{C}_\ell$, this
algorithm results in an ``optimised'' primordial spectrum
$\mathcal{P_R}^{\!\!\!\!\rm RL}(k)$, and is a very powerful tool to
find potential features in the spectrum.  We take $\mathcal{L}_{\rm
max}^{\rm X}$ to be the likelihood of the data for the primordial
spectrum $\mathcal{P_R}^{\!\!\!\!\rm RL}(k)$ combined with the
best-fit transfer function of the vanilla model.

This method does have two limitations though: as long as the transfer
function is held fixed, one is only sensitive to features that are not
degenerate with the cosmological parameters that determine the
transfer function (e.g., typically sharp features such as spikes or
oscillations, but not broad distortions).  It would be possible to
circumvent this problem by also optimising the transfer function, but
such a procedure would become computationally prohibitively expensive
in the context of our analysis.  Also, since we deconvolve binned
angular power spectrum data, we are not sensitive to extremely
high-frequency modulations of the primordial spectrum as reported in
Ref.~\cite{Ichiki:2009xs}.  Nevertheless, the $\Delta \chi^2_{\rm
  eff}$ obtained
%with the fixed transfer function
with our method provides a good measure of the potential deviation
from smoothness for realistic primordial spectra.

\subsection{Numerical implementation \label{sec:numimp}}

Applied to real data, the calculation of $\Delta \chi^2_{\rm eff}$
involves a maximisation of the likelihood using the reduced (i.e.,
$\tau = 0.089$) vanilla model, and the application of the
RL-deconvolution algorithm with the best-fit parameter transfer
function.

Naturally, the simulated data sets should be treated in exactly the
same way as the real data. Schematically, the procedure to be
performed for each random realisation of WMAP data should hence read
as follows:

\begin{itemize}
	\item[(i)]{Generate a random realisation
	$^{(i)}\mathcal{C}_\ell^{\rm sim}$ of simulated WMAP
	temperature data.}

	\item[(ii)]{Find minimal $\chi^2_{\rm eff,V}$ for a fit of the
	vanilla model to $^{(i)}\mathcal{C}_\ell^{\rm sim}$.}

	\item[{(iii)}]{Calculate the transfer function $^{(i)}T_\ell
	(k)$ for the best-fit parameters found in step (ii).}

	\item[(iv)]{Apply the modified Richardson-Lucy deconvolution
	algorithm to $^{(i)}\mathcal{C}_\ell^{\rm sim}$, using the
	transfer function $^{(i)}T_\ell (k)$.}

	\item[(v)]{Determine $\chi^2_{\rm eff,RL}$ from the deconvolved
	primordial spectrum and calculate the improvement $\Delta
	\chi^2_{\rm eff}$ over the power-law fit.}
\end{itemize}
To keep the task of simulating mock WMAP data at the level of
generating angular power spectra $^{(i)}\mathcal{C}_\ell^{\rm sim}$
and avoid having to generate random maps, we adopt a slightly
simplified form of the WMAP likelihood function (described in
Appendix~\ref{app:like}) throughout this paper.  For details of
simulating mock WMAP data (step (i)) we refer the reader to
Appendix~\ref{app:data}. 

The minimisation of step (ii) turns out to be the computationally most
expensive part of the algorithm.  Deterministic, simplex-like
minimisation routines like Numerical Recipes' \texttt{amoeba}
\cite{nr} are not very reliable for our purpose since they run the
risk of getting stuck in local minima.  Random-walk based routines,
such as simulated annealing are more suitable here.  We choose a
combination of the \texttt{amebsa}~\cite{nr} routine to find a good
starting point for a subsequent low-temperature run of the
Metropolis-Hastings algorithm (based on the \texttt{CosmoMC}
code~\cite{Lewis:2002ah}), which results in an estimate of the minimal
$\chi^2_{\rm eff,V}$ that is accurate to about 0.1.  Nonetheless, this
algorithm requires several thousand evaluations of CMB angular power
spectra and the corresponding likelihoods, which would take a
considerable time if one were to use a conventional Boltzmann code,
such as \texttt{CAMB}~\cite{Lewis:1999bs}.  We resort instead to the
interpolation code \texttt{PICO}~\cite{Fendt:2006uh,Fendt:2007uu} for
calculating the $\mathcal{C}_\ell$s, which can considerably speed up
this task, returning a reliable estimate for the minimal $\chi^2_{\rm
eff,V}$ within a few minutes on a regular desktop CPU.

Having found a good estimate of the maximum likelihood point, we then
use \texttt{CAMB} to calculate the transfer function for the best-fit
cosmological parameters in step (iii).  Having applied the
RL-deconvolution (step (iv)), giving us the optimised spectrum
$\mathcal{P_R}^{\!\!\!\!\rm RL}(k)$, we can finally compute the
improvement $\Delta \chi^2_{\rm eff}$ the optimised spectrum would
yield over the best power-law one (step (v)).

%\newpage
%%%%%%%%%%%%%%%%%%%%%%%%%%%%%%%%%%%%%%%%%%%%%%%%%%%%%%%%%%%%%%%%%%%%%%
\section{Results}                        
\label{sec:results}
%%%%%%%%%%%%%%%%%%%%%%%%%%%%%%%%%%%%%%%%%%%%%%%%%%%%%%%%%%%%%%%%%%%%%%

\subsection{Real data}

We commence by applying the RL-deconvolution method to the observed
WMAP5 temperature data.  The resulting best fit ``optimised''
primordial spectrum is shown in Fig.~\ref{fig:rlspectrum}; its
dominant features are qualitatively similar to those found in
Ref.~\cite{Shafieloo:2007tk} for the 3-year WMAP data: a cutoff-like
suppression at the largest scales accompanied by two prominent wiggles
around $k \sim 0.002\ {\rm Mpc}^{-1}$ and $k \sim 0.07\ {\rm
Mpc}^{-1}$.  The optimised spectrum gives $\chi^2_{\rm eff,RL} =
1029.45$, which marks an improvement of $\Delta \chi^2 = 24.41$ over
the power-law best fit of $\chi^2_{\rm eff,V} = 1053.86$.

\FIGURE{
\includegraphics[width=.66\textwidth,angle=0]{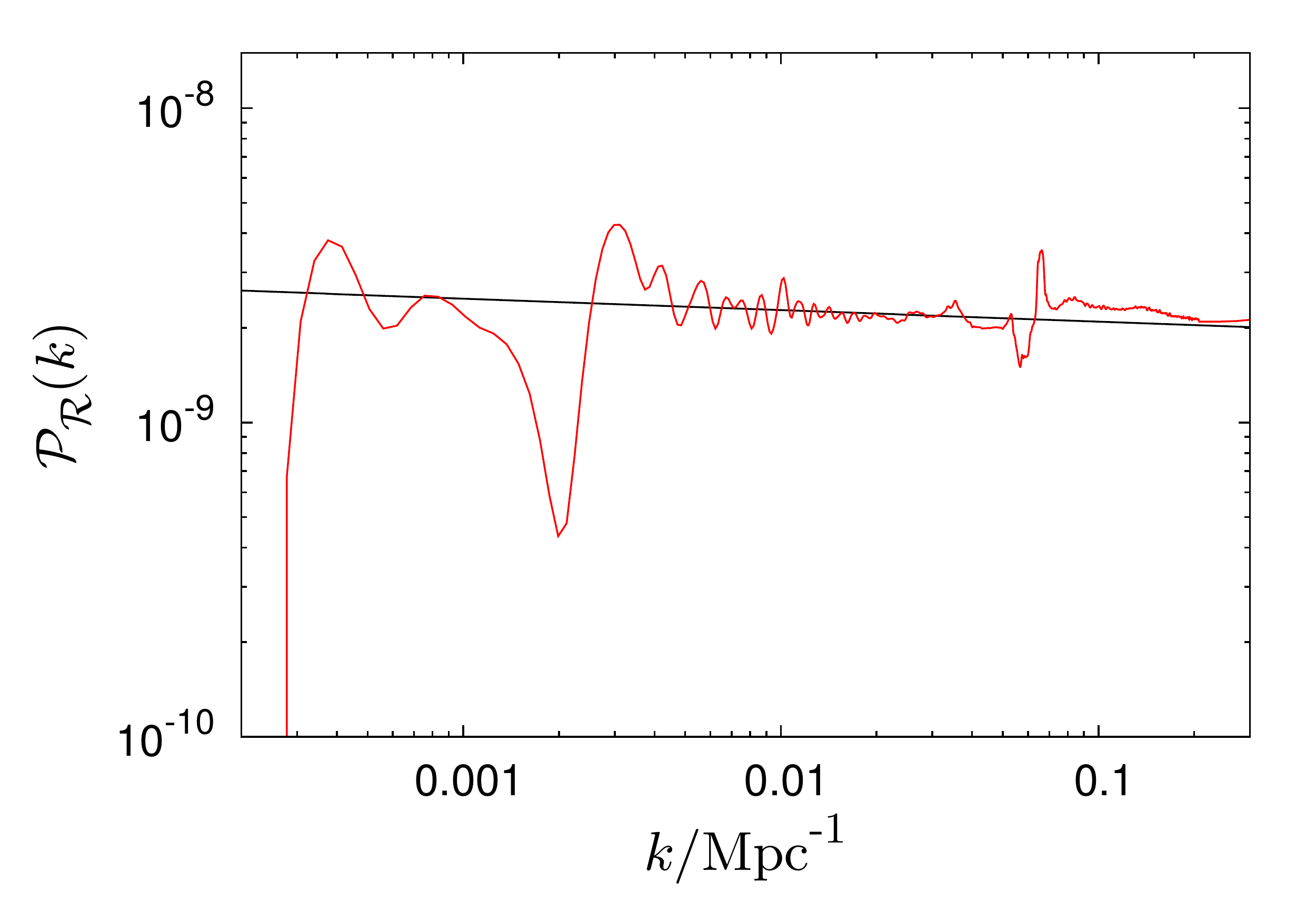} 
\caption{Best-fit power-law (black line), and RL-optimised primordial 
spectrum (red line). \label{fig:rlspectrum}}}

\subsection{Simulated data}

We have generated 2000 random realisations of simulated WMAP5
temperature data for multipoles $2 \leq \ell \leq 1000$ and applied
the procedure of Section~\ref{sec:numimp} to determine the $\Delta
\chi^2_{\rm eff}$ for each of them.  A histogram of the distribution
of $\Delta \chi^2_{\rm eff}$ values is shown in
Figure~{\ref{fig:histo}}.  Since the RL-algorithm is applied to binned
data whereas $\Delta \chi^2_{\rm eff}$ is defined using un-binned data,
for some simulated spectra the resulting $\Delta \chi^2_{\rm eff}$ can
be negative, even after wavelet smoothing.  In other words, in these
cases the RL-optimised spectra yield no improvement over a power-law,
and we set $\Delta \chi^2_{\rm eff} = 0$.  This has occurred in about
5\% of our simulated spectra and is marked by the red part of the
lowest-$\Delta \chi^2_{\rm eff}$-bin in the histogram.

\FIGURE{
\includegraphics[height=.66\textwidth,angle=270]{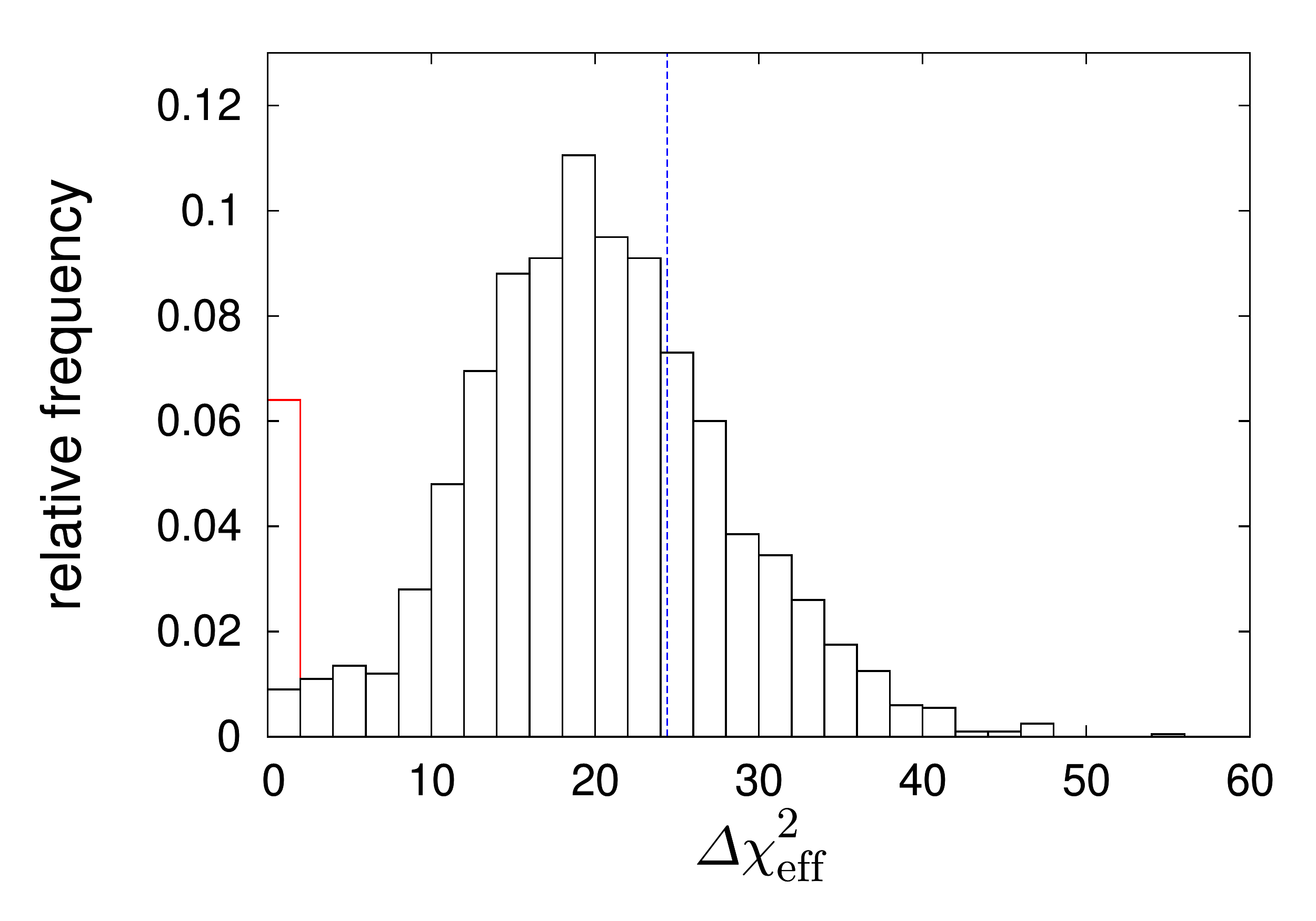}
\caption{Histogram showing the relative frequency of $\Delta
\chi^2_{\rm eff}$ for 2000 simulated realisations of the WMAP5
temperature spectrum. For about 5\% of the spectra, the RL-algorithm
does not result in a better fit; these cases are displayed in red.
The dotted blue line denotes the observed value $\Delta \chi^2_{\rm
eff} = 24.41$.  \label{fig:histo}}}

The vertical blue line in Figure~{\ref{fig:histo}} marks the observed
value $\Delta \chi^2_{\rm eff} = 24.41$.  Of the 2000 mock data sets,
525 have $\Delta \chi^2_{\rm eff} > 24.41$, corresponding to a
$p$-value of $\sim 0.26$.  To ensure this number is not affected by
our simplified likelihood-approximation, we calculated the observed
$\Delta\chi^2_{\rm eff}$ using the full WMAP5 likelihood code (version
\texttt{v3p2}), and found a difference of $\sim 1$.

Additionally, we considered statistics which are less susceptible to
low multipoles where the likelihood approximation is most inaccurate:
$\Delta \chi^2_{\rm eff}$ evaluated between $\ell_{\rm min}$ and
$\ell_{\max} = 1000$ with $\ell_{\rm min} = 10,\ 20,$ and $30$.  The
resulting $p$-values are 0.33, 0.14, and 0.31, respectively.  This
behaviour can be understood from the features observed in the real
data: removing $\ell < 10$ gets rid of a feature, leading to an
increase in the $p$-value; taking away $10 \leq \ell < 20$ removes a
part of the data which does not show features, causing the $p$-value
to drop, only to go up again when one removes the feature at $20 \leq
\ell < 30$.

\FIGURE{
\includegraphics[width=.66\textwidth,angle=0]{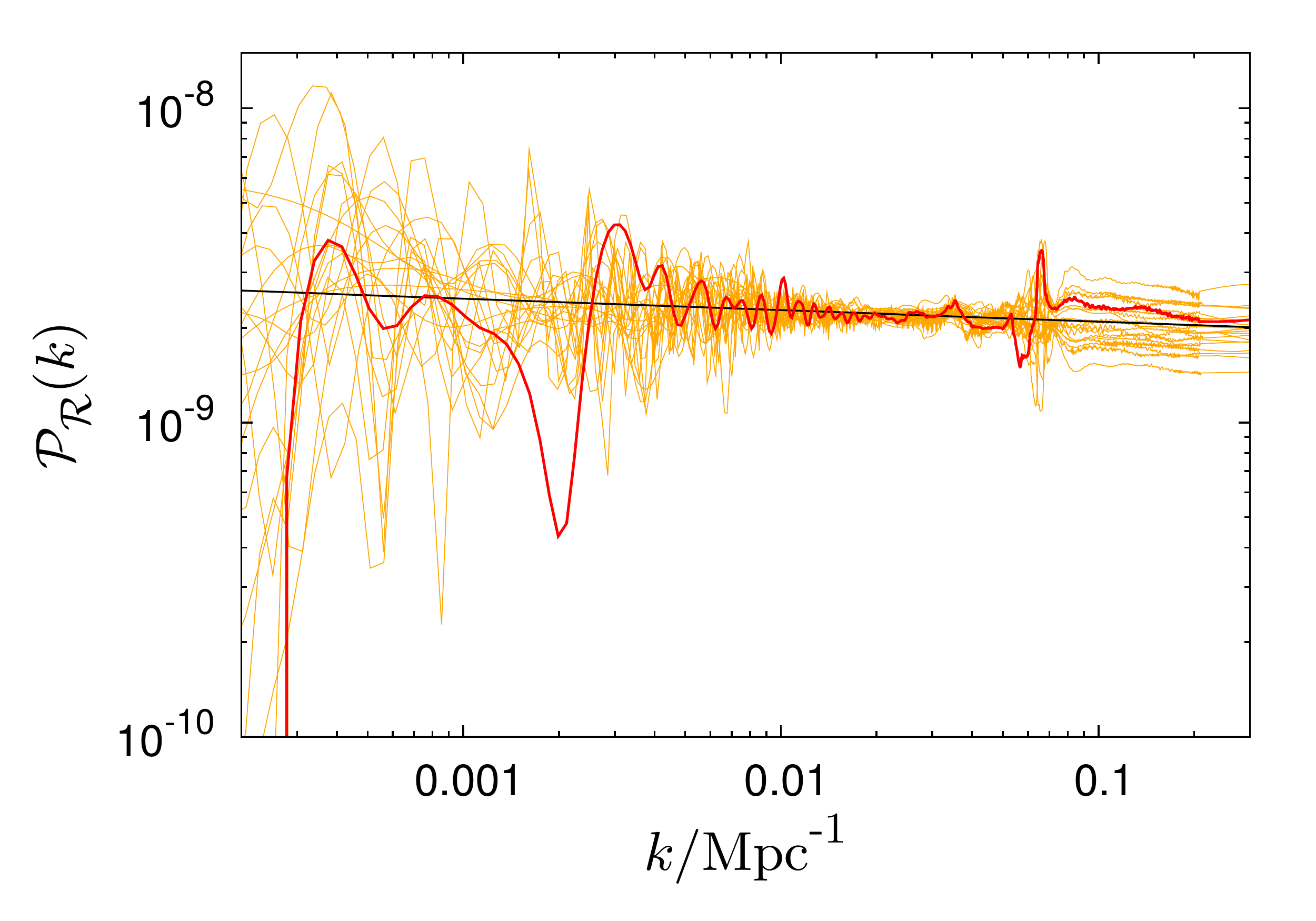}
\caption{Reconstructed primordial spectra for twenty mock data sets
  (orange lines).  For comparison we show the primordial spectrum
  reconstructed from WMAP5 data (red line) and our fiducial power-law
  spectrum (black line).  It is evident that the spectrum
  reconstructed from real data does not have an unusual amount of
  features.  The apparent feature at $0.05\ \rm{ Mpc}^{-1} < k < 0.07\
  \rm{ Mpc}^{-1}$ is caused by the noise term becoming dominant at the
  corresponding multipoles in the binned WMAP data.  For the last bin,
  $951 \le \ell \le 1000$, the averaged transfer function peaks at $k
  \sim 0.07\ \rm{ Mpc}^{-1}$. Beyond that the reconstructed primordial
  spectrum is no longer dominated by the peak of one data bin, but
  rather by the tails of the transfer functions of the last few bins;
  more structure here does not further improve the fit.
\label{fig:comparison}}}

In summary, the improvement in the fit from using an ``optimised''
primordial spectrum over a power-law spectrum is larger than the
observed one in roughly 30\% of our simulated data sets which assume
an underlying smooth spectrum. We thus conclude that the null
hypothesis cannot be rejected at high significance within the
limitations of our chosen statistic, and the data are compatible with
a smooth power-law primordial spectrum.  We illustrate this
qualitatively in Fig.~\ref{fig:comparison}, which shows that the
reconstructed spectrum from the observed data does not have an unusual
amount of ``featureness'' compared to the simulated data.

%\newpage

%%%%%%%%%%%%%%%%%%%%%%%%%%%%%%%%%%%%%%%%%%%%%%%%%%%%%%%%%%%%%%%%%%%%%%
\section{Conclusions}                        
\label{sec:conclusion}
%%%%%%%%%%%%%%%%%%%%%%%%%%%%%%%%%%%%%%%%%%%%%%%%%%%%%%%%%%%%%%%%%%%%%%

We have addressed the question whether the 5-year WMAP temperature
anisotropy data are compatible with the assumption of an underlying
smooth power-law primordial spectrum of curvature perturbations, or
whether they show any indication for features or other unaccounted-for
systematic effects. 
Assuming the true underlying primordial spectrum to be of power-law
form, we generated 2000 simulated WMAP angular temperature power
spectra, and estimated the amount of features by looking at the
improvement to the likelihood gained from fitting an ``optimised''
primordial spectrum, obtained by deconvolving the data,
instead of the usual power-law.  We found that 26\% of all simulated
data sets show a greater improvement than the one observed, which
leads us to conclude that the features seen in the WMAP 5-year
temperature data are not at all incompatible with the assumption of a
smooth primordial spectrum.\footnote{While this work was being
  completed, Ref.~\cite{Peiris:2009wp} appeared, which reaches similar
  conclusions from cross-validating a spline-reconstruction of the
  primordial spectrum from various cosmological data sets.}

However, we emphasise that our analysis does not disprove the
existence of features -- they are merely not strictly required by
present data.  Specifically, `simple' theoretical models that predict
similar features could conceivably remain favoured over a power-law.
Additionally, since the spectrum reconstruction method used here is
not sensitive to features below the binning scale of the WMAP data, we
cannot rule out the possibility of extremely high-frequency features.
We have demonstrated the feasibility of a full-scale frequentist
hypothesis testing analysis in the search for deviations from smooth
spectra though, and our method can just as well be applied to other
statistics which allow for even more general shapes of the primordial
spectrum.

Future analyses and new measurements may well reveal evidence for the
existence of features, and a search for them will certainly remain a
worthwhile endeavour, considering that any detection would profoundly
impact our understanding of the physics of inflation and also precise
estimation of the cosmological parameters~\cite{shafieloo09}.

Besides the imminent improvements to temperature anisotropy data from
the Planck satellite \cite{planck}, better measurements of the
$E$-polarisation of the CMB will greatly enhance sensitivity to
primordial features \cite{Nicholson:2009pi,Mortonson:2009qv} -- a
dedicated mission like CMBPol \cite{Baumann:2008aq} may prove
particularly helpful here.

%%%%%%%%%%%%%%%%%%%%%%%%%%%%%%%%%%%%%%%%%%%%%%%%%%%%%%%%%%%%%%%%%%%%%%
% Acknowledgments %%%%%%%%%%%%%%%%%%%%%%%%%%%%%%%%%%%%%%%%%%%%%%%%%%%%
%%%%%%%%%%%%%%%%%%%%%%%%%%%%%%%%%%%%%%%%%%%%%%%%%%%%%%%%%%%%%%%%%%%%%%
\acknowledgments{We thank Pedro Ferreira, Subir Sarkar and Yvonne Wong
for interesting discussions and comments. JH thanks the Oxford
Physics group for their hospitality during the initial stages of this
work.  Numerical work was performed on the MUST cluster at LAPP (CNRS
\& Universit\'e de Savoie).  JH acknowledges the support of a Feodor
Lynen-fellowship of the Alexander von Humboldt
foundation. AS acknowledges the support of the European Research
and Training Network MRTPNCT-2006 035863-1 (UniverseNet). }

\appendix

%%%%%%%%%%%%%%%%%%%%%%%%%%%%%%%%%%%%%%%%%%%%%%%%%%%%%%%%%%%%%%%%%%%%%%
\section{Appendix: Technical details} 
\label{sec:appendix}
%%%%%%%%%%%%%%%%%%%%%%%%%%%%%%%%%%%%%%%%%%%%%%%%%%%%%%%%%%%%%%%%%%%%%%

\subsection{Likelihood function\label{app:like}}

For reasons of simplicity we adopt here a form of the likelihood
function similar to the one used by the WMAP team in their first data
release~\cite{Verde:2003ey}.  For an input theoretical angular power
spectrum $\mathcal{C}^{\rm th}_\ell$, the likelihood of the data reads
\begin{equation}
	\chi^2_{\rm eff} \equiv -2 \ln \mathcal{L} = -2 \left(
	\frac{1}{3} \; \ln \mathcal{L}_{\rm Gauss} + \frac{2}{3} \;
	\ln \mathcal{L}'_{\rm LN} \right),
\end{equation}
with a Gaussian part
\begin{equation}
	-2 \ln \mathcal{L}_{\rm Gauss} = \sum_{\ell \ell'} (
	\mathcal{C}^{\rm th}_\ell - \hat{\mathcal{C}}_\ell )
	Q_{\ell \ell'} ( \mathcal{C}^{\rm th}_{\ell'} -
	\hat{\mathcal{C}}_{\ell'} ),  
\end{equation}
and a log-normal part
\begin{equation}
	-2 \ln \mathcal{L}'_{\rm LN} = \sum_{\ell \ell'} (z^{\rm
	th}_\ell - \hat{z}_\ell ) ( \mathcal{C}^{\rm
	th}_\ell + \mathcal{N}_\ell ) Q_{\ell \ell'} (
	\mathcal{C}^{\rm th}_{\ell'} + \mathcal{N}_{\ell'} )
	(z^{\rm th}_{\ell'} - \hat{z}_{\ell'} ),
\end{equation}
where $z^{\rm th}_\ell = \ln ( \mathcal{C}^{\rm th}_\ell +
\mathcal{N}_\ell )$ and $\hat{z}_\ell = \ln ( \hat{\mathcal{C}}_\ell +
\mathcal{N}_\ell )$, with $\hat{\mathcal{C}}_\ell$ the angular power
spectrum estimated from observation. The curvature matrix $Q_{\ell
\ell'}$ is given by
\begin{equation}
	\label{eq:curvmat}
	Q_{\ell \ell'} = D_\ell^{-1} \delta_{\ell \ell'} +
	\frac{r_{\ell \ell'}}{\sqrt{D_\ell D_{\ell'}}},
\end{equation}
with
\begin{equation}
	D_\ell = 2 \; \frac{ ( \mathcal{C}^{\rm th}_\ell +
	\mathcal{N}_\ell )^2}{( 2\ell +1 ) f_{\rm sky}^2}.
\end{equation}
The off-diagonal terms induced by the sky cut $r_{\ell \ell'}$, the
effective noise power spectrum $\mathcal{N}_\ell$ and the effective
sky fraction $f_{\rm sky}$ are supplied by the WMAP team
\cite{Nolta:2008ih} and available for download on the {\sc lambda}
web site\footnote{\texttt{http://lambda.gsfc.nasa.gov/}}.

\subsection{RL-algorithm for reconstructing the primordial spectrum
\label{app:rl}}

The Richardson-Lucy (RL) algorithm was developed and is widely used in
the context of image reconstruction in
astronomy~\cite{lucy74,rich72}. However, the method has also been
successfully used in cosmology, to deproject the $3$-D correlation
function and power spectrum from the measured $2$-D angular
correlation and $2$-D power spectrum~\cite{baug_efs93,baug_efs94}.

The angular power spectrum, $\mathcal{C}_\ell$, is a convolution of
the initial power spectrum $\mathcal{P_R}(k)$ generated in the early
universe, with a radiative transfer kernel $T_\ell (k)$ (the transfer
function), that is determined by the values of the cosmological
parameters. In our application, we solve the inverse problem of
determining the primordial power spectrum, $\mathcal{P_R}(k)$, from
the measured angular power spectrum, $\mathcal{C}_\ell$, using the
relation

\be 
\mathcal{C}_\ell\,= \int {\rm d}k \;  T_{\ell}(k) \mathcal{P_R}(k)
\simeq \sum_i T_{\ell} (k_i)\,\mathcal{P_R}(k_i). 
\label{clsum}
\ee

In the above equation, the {\em target} measured angular power
spectrum, $\mathcal{C}_\ell \equiv \mathcal{C}_\ell^{\rm obs}$, is the
data given by observations, and the radiative transport kernel,
\be
T_\ell(k_i) = \frac{\Delta k_i}{k_i}\,|{\Delta_{T\ell}(k_i,\eta_0)}|^2\,,
\label{glk} 
\ee 
encodes the response of the present multipoles of the
CMB perturbed photon distribution function $\Delta_{T \ell}(k_i,\eta_0)$
to unit of power per logarithm interval of wavenumber, $k$, in the
primordial perturbation spectrum. The kernel $T_\ell(k)$ is completely
fixed by the cosmological parameters of the {\em base} cosmological
model. The kernel $T_\ell(k)$ also includes the effect of geometrical
projection from the three-dimensional wavenumber, $k$, to the harmonic
multipole, $l$ on the two-dimensional sphere.

Obtaining $\mathcal{P_R}(k)$ from the measured $\mathcal{C}_\ell$ for
a given $T_\ell(k)$ is clearly a deconvolution problem.  An important
feature of the problem is that $\mathcal{C}_\ell^{\rm obs}$,
$T_\ell(k)$ and $\mathcal{P_R}(k)$ are all positive definite. We
employ an improved RL method to solve the inverse problem for
$\mathcal{P_R}(k)$ in Eq.~(\ref{clsum}).  The advantage of the RL
method is that positivity of the recovered $\mathcal{P_R}(k)$ is
automatically ensured, given positive definite $T_\ell(k)$ and
$\mathcal{C}_\ell$s. The RL method, readily derived from elementary
probability theory of distributions~\cite{lucy74}, is an iterative
method that can be neatly encoded into a simple recurrence
relation. The power spectrum $\mathcal{P_R}^{(i+1)}(k)$ recovered
after iteration $(i+1)$ is given by

\be 
\mathcal{P_R}^{(i+1)}(k)- \mathcal{P_R}^{(i)}(k)\,=\,\mathcal{P_R}^{(i)}(k)\,\sum_\ell\,\tilde T_\ell(k)\,
\zeta_k \,\frac {\tilde \mathcal{C}_\ell^{\rm obs}\,-\mathcal{C}_\ell^{(i)}}{\mathcal{C}_\ell^{(i)}}\,\,\,\tanh^2\,
\left[\frac{(\tilde \mathcal{C}^{\rm obs}_\ell\,-\mathcal{C}_\ell^{(i)})^2}{{\tilde \sigma_l}^2}\right],
\label{RLerr2} 
\ee
where $\tilde T_\ell(k)$ is the normalised kernel (on the $\ell$ space
for all $k$ wavenumbers), $\tilde \mathcal{C}^{\rm obs}_\ell$ is the
normalised measured data (target) and $\mathcal{C}_\ell^{(i)}$ is the
angular power spectrum at the $i^{\rm th}$ iteration obtained from
$\mathcal{C}_\ell^{(i)}\,= \sum \tilde
T_\ell(k)\,\mathcal{P_R}^{(i)}(k)$ using the recovered power spectrum
$\mathcal{P_R}^{(i)}(k)$. It is important to keep in mind that, due to
the formulation in terms of conditional probability distributions, the
RL method requires the kernel $T_\ell(k)$, data $\mathcal{C}_\ell$,
and the target vector $\mathcal{P_R}(k)$, all to be normalised at the
beginning,

\be
\sum_\ell\, \tilde \mathcal{C}_\ell\,=\,1; \quad \,\sum_k\, \tilde \mathcal{P_R}^{(1)}(k)\,=\,1;
\quad \,\sum_\ell\, \tilde T_\ell(k)\,=\,1, 
\ee 
where $\tilde \mathcal{P_R}^{(1)}(k)$ is the normalised initial guess
model of the primordial spectrum and the normalisation factor
$\zeta_k$ is defined by $\zeta_k \,=\, \sum_\ell\, T_\ell(k)\,$.

Note that Eq.(\ref{RLerr2}) represents a modified form of the
Richardson-Lucy algorithm specifically tailored for our purpose in
this problem~\cite{Shafieloo:2003gf,Shafieloo:2006hs,Shafieloo:2007tk}.

Following
Refs.~\cite{Shafieloo:2003gf,Shafieloo:2006hs,Shafieloo:2007tk}, we
apply the RL-algorithm to a {\em binned} version of the (real and
simulated) WMAP5 data.  The resulting raw deconvolved spectrum has
spurious oscillations and features on scales smaller than the bin
size, arising largely due to the $k$ space sampling and binning in
$\ell$ space, which adversely impact the (unbinned) full likelihood of
the data given the reconstructed spectrum.  As shown in
Ref.~\cite{Shafieloo:2006hs}, a subsequent smoothing of the raw
reconstructed spectrum with a discrete wavelet transform can lead to a
significant improvement in the likelihood.

\subsubsection{Discrete Wavelet Transform}

Wavelet transforms provide a powerful tool for the analysis of
transient and non-stationary data and is particularly useful in
picking out characteristic variations at different resolutions or
scales. This linear transform separates a data set in the form of
%low-pass or average coefficients, resembling the data itself, and
low-pass or average coefficients, which reflect the average behaviour
of the data, and wavelet or high-pass coefficients at different
levels, which capture the variations at corresponding scales. As
compared to Fourier or window Fourier transform, wavelets allow
optimal ``time-frequency'' localisation in the real as well as in the
Fourier domain. The vocabulary of DWT stems from applications in one
dimensional time-stream signal trains, but has found wide application
in signal in other domains and dimensions. Specifically in our case,
the signal being transformed is the primordial power spectrum,
$\mathcal{P_R}(k)$, a one dimensional function of wavenumber, $k$.

Wavelets are an orthonormal basis of small waves, with their
variations primarily concentrated in a finite region, which make them
ideal for analysing localised transient signals. Wavelets can be
continuous or discrete. In the latter case, the basis elements are
strictly finite in size, enabling them to achieve localisation, while
disentangling characteristic variations at different frequencies
\cite{Daubechies}.  For more details about DWT and its 
theoretical basis, see \cite{Shafieloo:2006hs}.

In this paper, we use the prescription of
Ref.~\cite{Shafieloo:2007tk}, and smooth the raw deconvolved spectrum
with the following method:
\begin{itemize}
	\item[1.]{Increase the number of $k$-values of the spectrum to a
power of 2, by padding evenly on both sides} 
	\item[2.]{Perform discrete wavelet transform}
	\item[3.]{Keep the $2^N$ wavelet coefficients corresponding to
lowest frequency features}
	\item[4.]{Perform inverse discrete wavelet transform and
calculate the likelihood for the smoothed spectrum}
	\item[5.]{Repeat steps 3.-4. for $N \in \left\{ 0, 11
\right\}$ and keep the spectrum that provides the best likelihood.} 
\end{itemize}

% The main goal is to reconstruct the primordial spectrum
%which lead to an angular power spectrum with a high likelihood to the
%entire $\mathcal{C}_{\ell}$ data at each multipole including the
%covariance between them. The WMAP likelihood of the
%$\mathcal{C}_{\ell}$ suffers owing to the spurious oscillations in the
%spectra on scales smaller than $\ell$ multipole space bins. 

% The WMAP
%likelihood improves as the spectra, $P(k)$ is smoothed. We use DWT to
%smooth the recovered spectrum so that WMAP likelihood of the
%corresponding theoretical $\mathcal{C}_\ell$ is maximised. Strictly
%speaking, the best likelihood obtained in our method is a lower bound
%leaving open a mathematical possibility of obtaining a superior
%likelihood at the given point of the cosmological parameter space with
%a different primordial power spectrum \footnote {There are some
%indications that the RL method can be related to a zero-noise limit of
%ML estimation.}.  Although it is difficult to establish that the final
%result is the unique solution with maximum likelihood, but numerous
%variations we have explored does suggest that it is perhaps very close
%to the best possible result. So we can claim that the improved
%reconstructed likelihoods which we drove for different points in the
%parameter space by assuming a broken scale invariant form of the
%primordial spectrum, put an upper limits for the best possible
%results~\cite{Shafieloo:2006hs,Shafieloo:2007tk}.

\subsection{Generating random realisations of WMAP data\label{app:data}}

Given an observed CMB temperature anisotropy $\Delta T(\hat{\mathbf
n})/\bar{T}$, after performing the usual expansion into spherical
harmonics
\begin{equation}
	\Delta T(\hat{\mathbf n})/\bar{T} = \sum_{\ell m} a^{\rm
	obs}_{\ell m} Y_{\ell m}(\hat{\mathbf n}),
\end{equation}
we can define an observed angular power spectrum
\begin{equation}
\label{eq:obspowspec}
	\mathfrak{C}^{\rm obs}_\ell \equiv \mathcal{C}^{\rm obs}_\ell
	+ \mathcal{N}^{\rm obs}_\ell = \frac{1}{2 \ell + 1}
	\sum_{m=-\ell}^\ell \left| a^{\rm obs}_{\ell m} \right|^2,
\end{equation}
which can be split up into an original signal $\mathcal{C}^{\rm
obs}_\ell$ and a contribution from experimental noise
$\mathcal{N}^{\rm obs}_\ell$.  Theory, on the other hand, can only
predict the average of this quantity over an ensemble of independent
observations, $\mathfrak{C}^{\rm th}_\ell = \langle \mathfrak{C}^{\rm
obs}_\ell\rangle$.  

In the following, we shall describe how to generate independent
realisations of simulated mock data (which replace the
$\hat{\mathcal{C}}_\ell$ in the associated mock likelihood function)
from a theoretical input spectrum $\mathfrak{C}^{\rm fid}_\ell$.  Since
the $\hat{\mathcal{C}}_\ell$ are actually defined as the difference
between the observed spectrum and the expected noise power spectrum,
$\hat{\mathcal{C}}_\ell =
\mathfrak{C}^{\rm obs}_\ell - \mathcal{N}^{\rm th}_\ell$, the
corresponding simulated quantity is $\mathfrak{C}^{\rm sim}_\ell -
\mathcal{N}^{\rm th}_\ell$.

For the sake of clarity we commence with the case of an ideal (i.e.,
noise-free, full-sky) observation.

\subsubsection{No experimental noise, full sky-coverage, no correlations}

Even in the case of an ideal measurement of the CMB temperature
anisotropies ($\mathcal{N}_\ell = 0$), the observed angular power
spectrum would still be subject to cosmic variance.  Under the
assumption of Gaussian fluctuations, the coefficients $a_{\ell m}$ can
be considered Gaussian random variables for the purpose of simulating
data sets $\mathcal{C}^{\rm sim}_\ell$.  Isotropy dictates $\langle 
a_{\ell m} a^*_{\ell' m'} \rangle = \mathcal{C}^{\rm fid}_\ell
\delta_{\ell \ell'} \delta_{m m'}$, so different multipoles will be
independent of each other.  Hence, for each $\ell$ the
simulated $\mathcal{C}^{\rm sim}_\ell$ can be constructed by drawing
random numbers $x_\ell$ from a $\chi^2_{2\ell + 1}$-distribution:
\begin{equation}
	\mathfrak{C}^{\rm sim}_\ell = \mathcal{C}^{\rm sim}_\ell =
	\mathcal{C}^{\rm fid}_\ell \frac{x_\ell}{\langle x_\ell
	\rangle} = \mathcal{C}^{\rm fid}_\ell + \mathcal{C}^{\rm
	fid}_\ell \left( \frac{x_\ell}{2 \ell + 1} - 1 \right).
\end{equation}

\subsubsection{Adding experimental noise}

Realistic observations are of course subject to instrumental noise.

Assuming the noise to be isotropic and Gaussian, its effect on the
distribution the $a_{\ell m}$ are drawn from can be
considered as a convolution with a Gaussian of width
$\sqrt{\mathcal{N}^{\rm th}_\ell}$, and we have
\begin{equation}
	\mathfrak{C}^{\rm sim}_\ell - \mathcal{N}^{\rm th}_\ell =
	\mathcal{C}^{\rm sim}_\ell + \mathcal{N}^{\rm sim}_\ell -
	\mathcal{N}^{\rm th}_\ell = \mathcal{C}^{\rm fid}_\ell + \left(
	\mathcal{C}^{\rm fid}_\ell + \mathcal{N}^{\rm th}_\ell \right)
	\left( \frac{x_\ell}{2 \ell + 1} - 1 \right).
\end{equation}

\subsubsection{Partial sky coverage and correlations}

Even for nominally full-sky observations, some parts of the sky cannot
be used to construct spectra and need to be cut out, since the signal
is obscured by point sources and galactic foregrounds.  This has two
consequences on the recovered $\hat{\mathcal{C}}_\ell$.  First, using
only part of the sky leads to a loss of information, resulting in a
larger scatter of $\hat{\mathcal{C}}_\ell$ around the true underlying
spectrum $\mathcal{C}^{\rm fid}_\ell$.  Second, since the sky cut
breaks isotropy, the (pseudo)-$\hat{\mathcal{C}}_\ell$ constructed
from an incomplete map will no longer be
uncorrelated~\cite{Wandelt:2000av}.

Since the curvature matrix (\ref{eq:curvmat}) appearing in the
likelihood function scales like $f_{\rm sky}^2$, the standard
deviation of $\mathcal{C}_\ell^{\rm sim} - \mathcal{C}_\ell^{\rm fid}$
should scale like $f_{\rm sky}^{-1}$.  This increase in scatter around
the fiducial model can be modelled by drawing the random variables
$x_\ell$ from a $\chi_{f^2_{\rm sky}(2 \ell + 1)}^2$-distribution.
Since it is numerically much easier to generate $\chi^2$ variates with
an integer number of degrees of freedom we round $f^2_{\rm sky}(2 \ell
+ 1)$ to the nearest integer and eventually rescale the resulting
$y_\ell \equiv x_\ell/\langle x_\ell \rangle - 1$ accordingly, to
recover the correct standard deviation.

The so-generated $y_\ell$ are uncorrelated random variables.  For the
simulated spectrum to have the same correlation properties as the real
WMAP data, we take instead the vector
\begin{equation}
	(\tilde{y}_\ell) \equiv {\rm Chol}(K_{\ell \ell'}) (y_{\ell'}),
\end{equation}
where the lower triangular matrix ${\rm Chol}(K_{\ell \ell'})$ is the
Cholesky decomposition of the correlation matrix $K_{\ell \ell'} =
\delta_{\ell \ell'} + r_{\ell \ell'}$.  Finally, the simulated spectra
are given by
\begin{equation}
	\label{eq:chol}
	\mathfrak{C}^{\rm sim}_\ell - \mathcal{N}^{\rm th}_\ell =
	\mathcal{C}^{\rm fid}_\ell + \left( \mathcal{C}^{\rm fid}_\ell +
	\mathcal{N}^{\rm th}_\ell \right) \tilde{y}_\ell.
\end{equation}
The above procedure can rarely result in negative values of
$\mathcal{C}^{\rm sim}_\ell + \mathcal{N}_\ell^{\rm sim}$ at low
multipoles, due to the rescaling of the $y_\ell$ or due to the
transformation of Eq.~(\ref{eq:chol}).  Such data is unphysical and
not compatible with the log-normal piece of the likelihood function
and must therefore be discarded.  As a result, the simulated data can
be slightly biased; however, the bias is negligible and remains
smaller than $\mathcal{O}$(1\%) of cosmic variance for all $\ell$.

%%%%%%%%%%%%%%%%%%%%%%%%%%%%%%%%%%%%%%%%%%%%%%%%%%%%%%%%%%%%%%%%%%%%%%
% References %%%%%%%%%%%%%%%%%%%%%%%%%%%%%%%%%%%%%%%%%%%%%%%%%%%%%%%%%
%%%%%%%%%%%%%%%%%%%%%%%%%%%%%%%%%%%%%%%%%%%%%%%%%%%%%%%%%%%%%%%%%%%%%%


\begin{thebibliography}{99}

\bibitem{Starobinsky:1982ee}
  A.~A.~Starobinsky,
  %``Dynamics Of Phase Transition In The New Inflationary Universe Scenario And
  %Generation Of Perturbations,''
  Phys.\ Lett.\  B {\bf 117} (1982) 175.
  %%CITATION = PHLTA,B117,175;%%

\bibitem{Guth:1982ec}
  A.~H.~Guth and S.~Y.~Pi,
  %``Fluctuations In The New Inflationary Universe,''
  Phys.\ Rev.\ Lett.\  {\bf 49} (1982) 1110.
  %%CITATION = PRLTA,49,1110;%%

\bibitem{Bardeen:1983qw}
  J.~M.~Bardeen, P.~J.~Steinhardt and M.~S.~Turner,
  %``Spontaneous Creation Of Almost Scale - Free Density Perturbations In An
  %Inflationary Universe,''
  Phys.\ Rev.\  D {\bf 28} (1983) 679.
  %%CITATION = PHRVA,D28,679;%%

\bibitem{Mukhanov:1990me}
  V.~F.~Mukhanov, H.~A.~Feldman and R.~H.~Brandenberger,
  %``Theory of cosmological perturbations. Part 1. Classical perturbations. Part
  %2. Quantum theory of perturbations. Part 3. Extensions,''
  Phys.\ Rept.\  {\bf 215} (1992) 203.
  %%CITATION = PRPLC,215,203;%%

\bibitem{Martin:1999fa}
  J.~Martin, A.~Riazuelo and M.~Sakellariadou,
  %``Non-vacuum initial states for cosmological perturbations of
  %quantum-mechanical origin,''
  Phys.\ Rev.\  D {\bf 61} (2000) 083518
  [arXiv:astro-ph/9904167].
  %%CITATION = PHRVA,D61,083518;%%

\bibitem{Martin:2000xs}
  J.~Martin and R.~H.~Brandenberger,
  %``The trans-Planckian problem of inflationary cosmology,''
  Phys.\ Rev.\  D {\bf 63} (2001) 123501
  [arXiv:hep-th/0005209].
  %%CITATION = PHRVA,D63,123501;%%

\bibitem{Danielsson:2002kx}
  U.~H.~Danielsson,
  %``A note on inflation and transplanckian physics,''
  Phys.\ Rev.\  D {\bf 66} (2002) 023511
  [arXiv:hep-th/0203198].
  %%CITATION = PHRVA,D66,023511;%%

\bibitem{Contaldi:2003zv}
  C.~R.~Contaldi, M.~Peloso, L.~Kofman and A.~Linde,
  %``Suppressing the lower Multipoles in the CMB Anisotropies,''
  JCAP {\bf 0307} (2003) 002
  [arXiv:astro-ph/0303636].
  %%CITATION = JCAPA,0307,002;%%

\bibitem{Adams:1997de}
  J.~A.~Adams, G.~G.~Ross and S.~Sarkar,
  %``Multiple inflation,''
  Nucl.\ Phys.\  B {\bf 503} (1997) 405
  [arXiv:hep-ph/9704286].
  %%CITATION = NUPHA,B503,405;%%

\bibitem{Joy:2007na}
  M.~Joy, V.~Sahni and A.~A.~Starobinsky,
  %``A New Universal Local Feature in the Inflationary Perturbation Spectrum,''
  Phys.\ Rev.\  D {\bf 77} (2008) 023514
  [arXiv:0711.1585 [astro-ph]].
  %%CITATION = PHRVA,D77,023514;%%

\bibitem{Starobinsky:1992ts}
  A.~A.~Starobinsky,
  %``Spectrum Of Adiabatic Perturbations In The Universe When There Are
  %Singularities In The Inflation Potential,''
  JETP Lett.\  {\bf 55} (1992) 489
  [Pisma Zh.\ Eksp.\ Teor.\ Fiz.\  {\bf 55} (1992) 477].
  %%CITATION = ZFPRA,55,477;%%

\bibitem{Adams:2001vc}
  J.~A.~Adams, B.~Cresswell and R.~Easther,
  %``Inflationary perturbations from a potential with a step,''
  Phys.\ Rev.\  D {\bf 64} (2001) 123514
  [arXiv:astro-ph/0102236].
  %%CITATION = PHRVA,D64,123514;%%

\bibitem{Jain:2008dw}
  R.~K.~Jain, P.~Chingangbam, J.~O.~Gong, L.~Sriramkumar and T.~Souradeep,
  %``Double inflation and the low CMB multipoles,''
  JCAP {\bf 0901} (2009) 009
  [arXiv:0809.3915 [astro-ph]].
  %%CITATION = JCAPA,0901,009;%%

\bibitem{Chung:1999ve}
  D.~J.~H.~Chung, E.~W.~Kolb, A.~Riotto and I.~I.~Tkachev,
  %``Probing Planckian physics: Resonant production of particles during
  %inflation and features in the primordial power spectrum,''
  Phys.\ Rev.\  D {\bf 62} (2000) 043508
  [arXiv:hep-ph/9910437].
  %%CITATION = PHRVA,D62,043508;%%

\bibitem{Elgaroy:2003hp}
  \O.~Elgar\o y, S.~Hannestad and T.~Haugb\o lle,
  %``Observational constraints on particle production during inflation,''
  JCAP {\bf 0309} (2003) 008
  [arXiv:astro-ph/0306229].
  %%CITATION = JCAPA,0309,008;%%

\bibitem{Kofman:1986wm}
  L.~A.~Kofman and A.~D.~Linde,
  %``Generation of Density Perturbations in the Inflationary Cosmology,''
  Nucl.\ Phys.\  B {\bf 282} (1987) 555.
  %%CITATION = NUPHA,B282,555;%%

\bibitem{KofPog1988}
  L.~A.~Kofman and D.~Y.~Pogosyan, 
  Phys. Lett. B {\bf 214}, 508 (1988).

\bibitem{Salopek:1988qh}
  D.~S.~Salopek, J.~R.~Bond and J.~M.~Bardeen,
  %``Designing Density Fluctuation Spectra in Inflation,''
  Phys.\ Rev.\  D {\bf 40} (1989) 1753.
  %%CITATION = PHRVA,D40,1753;%%

\bibitem{Polarski:1992dq}
  D.~Polarski and A.~A.~Starobinsky,
  %``Spectra of perturbations produced by double inflation with an intermediate
  %matter dominated stage,''
  Nucl.\ Phys.\  B {\bf 385} (1992) 623.
  %%CITATION = NUPHA,B385,623;%%

\bibitem{Kaloper:2003nv}
  N.~Kaloper and M.~Kaplinghat,
  %``Primeval corrections to the CMB anisotropies,''
  Phys.\ Rev.\  D {\bf 68} (2003) 123522
  [arXiv:hep-th/0307016].
  %%CITATION = PHRVA,D68,123522;%%

\bibitem{Cline:2006db}
  J.~M.~Cline and L.~Hoi,
  %``Inflationary potential reconstruction for a WMAP running power  spectrum,''
  JCAP {\bf 0606} (2006) 007
  [arXiv:astro-ph/0603403].
  %%CITATION = JCAPA,0606,007;%%

\bibitem{Jain:2009pm}
  R.~K.~Jain, P.~Chingangbam, L.~Sriramkumar and T.~Souradeep,
  %``The tensor-to-scalar ratio in punctuated inflation,''
  arXiv:0904.2518 [astro-ph.CO].
  %%CITATION = ARXIV:0904.2518;%%

\bibitem{Nolta:2008ih}
  M.~R.~Nolta {\it et al.}  [WMAP Collaboration],
  %``Five-Year Wilkinson Microwave Anisotropy Probe (WMAP) Observations: Angular
  %Power Spectra,''
  Astrophys.\ J.\ Suppl.\  {\bf 180} (2009) 296
  [arXiv:0803.0593 [astro-ph]].
  %%CITATION = APJSA,180,296;%%


\bibitem{Dunkley:2008ie}
  J.~Dunkley {\it et al.}  [WMAP Collaboration],
  %``Five-Year Wilkinson Microwave Anisotropy Probe (WMAP) Observations:
  %Likelihoods and Parameters from the WMAP data,''
  Astrophys.\ J.\ Suppl.\  {\bf 180} (2009) 306
  [arXiv:0803.0586 [astro-ph]].
  %%CITATION = APJSA,180,306;%%

\bibitem{Komatsu:2008hk}
  E.~Komatsu {\it et al.}  [WMAP Collaboration],
  %``Five-Year Wilkinson Microwave Anisotropy Probe (WMAP)
  %Observations:Cosmological Interpretation,''
  Astrophys.\ J.\ Suppl.\  {\bf 180} (2009) 330
  [arXiv:0803.0547 [astro-ph]].
  %%CITATION = APJSA,180,330;%%

\bibitem{Barriga:2000nk}
  J.~Barriga, E.~Gaztanaga, M.~G.~Santos and S.~Sarkar,
  %``On the APM power spectrum and the CMB anisotropy: Evidence for a phase
  %transition during inflation?,''
  Mon.\ Not.\ Roy.\ Astron.\ Soc.\  {\bf 324} (2001) 977
  [arXiv:astro-ph/0011398].
  %%CITATION = MNRAA,324,977;%%

\bibitem{Peiris:2003ff}
  H.~V.~Peiris {\it et al.}  [WMAP Collaboration],
  %``First year Wilkinson Microwave Anisotropy Probe (WMAP) observations:
  %Implications for inflation,''
  Astrophys.\ J.\ Suppl.\  {\bf 148} (2003) 213
  [arXiv:astro-ph/0302225].
  %%CITATION = APJSA,148,213;%%

\bibitem{Martin:2003sg}
  J.~Martin and C.~Ringeval,
  %``Superimposed Oscillations in the WMAP Data?,''
  Phys.\ Rev.\  D {\bf 69} (2004) 083515
  [arXiv:astro-ph/0310382].
  %%CITATION = PHRVA,D69,083515;%%

\bibitem{Martin:2004iv}
  J.~Martin and C.~Ringeval,
  %``Addendum to ``Superimposed Oscillations in the WMAP Data?'',''
  Phys.\ Rev.\  D {\bf 69} (2004) 127303
  [arXiv:astro-ph/0402609].
  %%CITATION = PHRVA,D69,127303;%%

\bibitem{Martin:2004yi}
  J.~Martin and C.~Ringeval,
  %``Exploring the superimposed oscillations parameter space,''
  JCAP {\bf 0501} (2005) 007
  [arXiv:hep-ph/0405249].
  %%CITATION = JCAPA,0501,007;%%

\bibitem{Hunt:2004vt}
 P.~Hunt and S.~Sarkar,
  %``Multiple inflation and the WMAP 'glitches',''
  Phys.\ Rev.\  D {\bf 70} (2004) 103518
 [arXiv:astro-ph/0408138].
  %%CITATION = PHRVA,D70,103518;%%

\bibitem{Hunt:2007dn}
  P.~Hunt and S.~Sarkar,
  %``Multiple inflation and the WMAP 'glitches' II. Data analysis and
  %cosmological parameter extraction,''
  Phys.\ Rev.\  D {\bf 76} (2007) 123504
  [arXiv:0706.2443 [astro-ph]].
  %%CITATION = PHRVA,D76,123504;%%

\bibitem{Joy:2008qd}
  M.~Joy, A.~Shafieloo, V.~Sahni and A.~A.~Starobinsky,
  %``Is a step in the primordial spectral index favored by CMB data ?,''
  JCAP {\bf 0906} (2009) 028
  [arXiv:0807.3334 [astro-ph]].
  %%CITATION = JCAPA,0906,028;%%

\bibitem{Bridges:2005br}
  M.~Bridges, A.~N.~Lasenby and M.~P.~Hobson,
  %``A Bayesian analysis of the primordial power spectrum,''
  Mon.\ Not.\ Roy.\ Astron.\ Soc.\  {\bf 369} (2006) 1123
  [arXiv:astro-ph/0511573].
  %%CITATION = MNRAA,369,1123;%%

\bibitem{Spergel:2006hy}
  D.~N.~Spergel {\it et al.}  [WMAP Collaboration],
  %``Wilkinson Microwave Anisotropy Probe (WMAP) three year results:
  %Implications for cosmology,''
  Astrophys.\ J.\ Suppl.\  {\bf 170} (2007) 377
  [arXiv:astro-ph/0603449].
  %%CITATION = APJSA,170,377;%%

\bibitem{Covi:2006ci}
  L.~Covi, J.~Hamann, A.~Melchiorri, A.~Slosar and I.~Sorbera,
  %``Inflation and WMAP three year data: Features have a future!,''
  Phys.\ Rev.\  D {\bf 74} (2006) 083509
  [arXiv:astro-ph/0606452].
  %%CITATION = PHRVA,D74,083509;%%

\bibitem{Hamann:2007pa}
  J.~Hamann, L.~Covi, A.~Melchiorri and A.~Slosar,
  %``New constraints on oscillations in the primordial spectrum of inflationary
  %perturbations,''
  Phys.\ Rev.\  D {\bf 76} (2007) 023503
  [arXiv:astro-ph/0701380].
  %%CITATION = PHRVA,D76,023503;%%

\bibitem{Barnaby:2009dd}
  N.~Barnaby and Z.~Huang,
  %``Particle Production During Inflation: Observational Constraints and
  %Signatures,''
  arXiv:0909.0751 [astro-ph.CO].
  %%CITATION = ARXIV:0909.0751;%%

\bibitem{Ichiki:2009xs}
  K.~Ichiki, R.~Nagata and J.~Yokoyama,
  %``Cosmic Discordance: Detection of a modulation in the primordial fluctuation
  %spectrum,''
  arXiv:0911.5108 [astro-ph.CO].
  %%CITATION = ARXIV:0911.5108;%%

\bibitem{Bridle:2003sa}
  S.~L.~Bridle, A.~M.~Lewis, J.~Weller and G.~Efstathiou,
  %``Reconstructing the primordial power spectrum,''
  Mon.\ Not.\ Roy.\ Astron.\ Soc.\  {\bf 342} (2003) L72
  [arXiv:astro-ph/0302306].
  %%CITATION = MNRAA,342,L72;%%

\bibitem{Hannestad:2003zs}
  S.~Hannestad,
  %``Reconstructing the primordial power spectrum - a new algorithm,''
  JCAP {\bf 0404} (2004) 002
  [arXiv:astro-ph/0311491].
  %%CITATION = JCAPA,0404,002;%%

\bibitem{Mukherjee:2003ag}
  P.~Mukherjee and Y.~Wang,
  %``Model-Independent Reconstruction of the Primordial Power Spectrum from WMAP
  %Data,''
  Astrophys.\ J.\  {\bf 599} (2003) 1
  [arXiv:astro-ph/0303211].
  %%CITATION = ASJOA,599,1;%%


\bibitem{Leach:2005av}
  S.~M.~Leach,
  %``Measuring the primordial power spectrum: Principal component analysis of
  %the cosmic microwave background,''
  Mon.\ Not.\ Roy.\ Astron.\ Soc.\  {\bf 372} (2006) 646
  [arXiv:astro-ph/0506390].
  %%CITATION = MNRAA,372,646;%%

\bibitem{Kogo:2003yb}
  N.~Kogo, M.~Matsumiya, M.~Sasaki and J.~Yokoyama,
  %``Reconstructing the primordial spectrum from WMAP data by the cosmic
  %inversion method,''
  Astrophys.\ J.\  {\bf 607} (2004) 32
  [arXiv:astro-ph/0309662].
  %%CITATION = ASJOA,607,32;%%

\bibitem{Shafieloo:2003gf}
  A.~Shafieloo and T.~Souradeep,
  %``Primordial power spectrum from WMAP,''
  Phys.\ Rev.\  D {\bf 70} (2004) 043523
  [arXiv:astro-ph/0312174].
  %%CITATION = PHRVA,D70,043523;%%

\bibitem{TocchiniValentini:2004ht}
  D.~Tocchini-Valentini, M.~Douspis and J.~Silk,
  %``Are there features in the primordial power spectrum?,''
  Mon.\ Not.\ Roy.\ Astron.\ Soc.\  {\bf 359} (2005) 31
  [arXiv:astro-ph/0402583].
  %%CITATION = MNRAA,359,31;%%

\bibitem{Shafieloo:2006hs}
  A.~Shafieloo, T.~Souradeep, P.~Manimaran, P.~K.~Panigrahi and R.~Rangarajan,
  %``Features in the Primordial Spectrum from WMAP: A Wavelet Analysis,''
  Phys.\ Rev.\  D {\bf 75} (2007) 123502
  [arXiv:astro-ph/0611352].
  %%CITATION = PHRVA,D75,123502;%%

\bibitem{Shafieloo:2007tk}
  A.~Shafieloo and T.~Souradeep,
  %``Estimation of Primordial Spectrum with post-WMAP 3 year data,''
  Phys.\ Rev.\  D {\bf 78} (2008) 023511
  [arXiv:0709.1944 [astro-ph]].
  %%CITATION = PHRVA,D78,023511;%%

\bibitem{Nagata:2008tk}
  R.~Nagata and J.~Yokoyama,
  %``Reconstruction of the primordial fluctuation spectrum from the five-year
  %WMAP data by the cosmic inversion method with band-power decorrelation
  %analysis,''
  Phys.\ Rev.\  D {\bf 78} (2008) 123002
  [arXiv:0809.4537 [astro-ph]].
  %%CITATION = PHRVA,D78,123002;%%

\bibitem{Nagata:2008zj}
  R.~Nagata and J.~Yokoyama,
  %``Band-power reconstruction of the primordial fluctuation spectrum by the
  %maximum likelihood reconstruction method,''
  Phys.\ Rev.\  D {\bf 79} (2009) 043010
  [arXiv:0812.4585 [astro-ph]].
  %%CITATION = PHRVA,D79,043010;%%

\bibitem{Ichiki:2009zz}
  K.~Ichiki and R.~Nagata,
  %``Brute force reconstruction of the primordial fluctuation spectrum from
  %five-year Wilkinson Microwave Anisotropy Probe observations,''
  Phys.\ Rev.\  D {\bf 80} (2009) 083002.
  %%CITATION = PHRVA,D80,083002;%%

\bibitem{Nicholson:2009pi}
  G.~Nicholson and C.~R.~Contaldi,
  %``Reconstruction of the Primordial Power Spectrum using Temperature and
  %Polarisation Data from Multiple Experiments,''
  arXiv:0903.1106 [astro-ph.CO].
  %%CITATION = ARXIV:0903.1106;%%

\bibitem{Nicholson:2009zj}
  G.~Nicholson, C.~R.~Contaldi and P.~Paykari,
  %``Reconstruction of the Primordial Power Spectrum by Direct Inversion,''
  arXiv:0909.5092 [astro-ph.CO].
  %%CITATION = ARXIV:0909.5092;%%

\bibitem{Trotta:2005ar}
  R.~Trotta,
  %``Applications of Bayesian model selection to cosmological parameters,''
  Mon.\ Not.\ Roy.\ Astron.\ Soc.\  {\bf 378} (2007) 72
  [arXiv:astro-ph/0504022].
  %%CITATION = MNRAA,378,72;%%

\bibitem{Liddle:2009xe}
  A.~R.~Liddle,
  %``Statistical methods for cosmological parameter selection and estimation,''
  arXiv:0903.4210 [hep-th].
  %%CITATION = ARXIV:0903.4210;%%

\bibitem{Efstathiou:2008ed}
  G.~Efstathiou,
  %``Limitations of Bayesian Evidence Applied to Cosmology,''
  arXiv:0802.3185 [astro-ph].
  %%CITATION = ARXIV:0802.3185;%%

\bibitem{Spergel:2003cb}
  D.~N.~Spergel {\it et al.}  [WMAP Collaboration],
  %``First Year Wilkinson Microwave Anisotropy Probe (WMAP) Observations:
  %Determination of Cosmological Parameters,''
  Astrophys.\ J.\ Suppl.\  {\bf 148} (2003) 175
  [arXiv:astro-ph/0302209].
  %%CITATION = APJSA,148,175;%%

\bibitem{Lewis:2003qm}
  A.~Lewis,
  %``Cosmological parameters and the WMAP data,''
  arXiv:astro-ph/0310186.
  %%CITATION = ASTRO-PH/0310186;%%

\bibitem{nr}
  W.~H.~Press, S.~A.~Teukolsky, W.~T.~Vetterling and B.~P.~Flannery,
  ``Numerical Recipes in Fortran,''
  Cambridge University Press, 1992

\bibitem{Lewis:2002ah}
  A.~Lewis and S.~Bridle,
  %``Cosmological parameters from CMB and other data:
  %A Monte-Carlo approach,''
  Phys.\ Rev.\ D {\bf 66} (2002) 103511
  [astro-ph/0205436].
  %%CITATION = ASTRO-PH 0205436;%%

\bibitem{Lewis:1999bs}
  A.~Lewis, A.~Challinor and A.~Lasenby,
  %``Efficient Computation of CMB anisotropies in closed FRW models,''
  Astrophys.\ J.\  {\bf 538} (2000) 473
  [arXiv:astro-ph/9911177].
  %%CITATION = ASJOA,538,473;%%

\bibitem{Fendt:2006uh}
  W.~A.~Fendt and B.~D.~Wandelt,
  %``Pico: Parameters for the Impatient Cosmologist,''
  Astrophys.\ J.\  {\bf 654} (2006) 2
  [arXiv:astro-ph/0606709].
  %%CITATION = ASJOA,654,2;%%

\bibitem{Fendt:2007uu}
  W.~A.~Fendt and B.~D.~Wandelt,
  %``Computing High Accuracy Power Spectra with Pico,''
  arXiv:0712.0194 [astro-ph].
  %%CITATION = ARXIV:0712.0194;%%

\bibitem{shafieloo09}
A.~Shafieloo and T.~Souradeep, arXive:0901.0716 [astro-ph]

\bibitem{Peiris:2009wp}
  H.~V.~Peiris and L.~Verde,
  %``The Shape of the Primordial Power Spectrum: A Last Stand Before Planck,''
  arXiv:0912.0268 [astro-ph.CO].
  %%CITATION = ARXIV:0912.0268;%%

\bibitem{planck}
    [Planck Collaboration],
  %``Planck: The scientific programme,''
  arXiv:astro-ph/0604069.
  %%CITATION = ASTRO-PH/0604069;%%

\bibitem{Mortonson:2009qv}
  M.~J.~Mortonson, C.~Dvorkin, H.~V.~Peiris and W.~Hu,
  %``CMB polarization features from inflation versus reionization,''
  Phys.\ Rev.\  D {\bf 79} (2009) 103519
  [arXiv:0903.4920 [astro-ph.CO]].
  %%CITATION = PHRVA,D79,103519;%%

\bibitem{Baumann:2008aq}
  D.~Baumann {\it et al.}  [CMBPol Study Team Collaboration],
  %``CMBPol Mission Concept Study: Probing Inflation with CMB Polarization,''
  AIP Conf.\ Proc.\  {\bf 1141} (2009) 10
  [arXiv:0811.3919 [astro-ph]].
  %%CITATION = APCPC,1141,10;%%

\bibitem{Verde:2003ey}
  L.~Verde {\it et al.}  [WMAP Collaboration],
  %``First Year Wilkinson Microwave Anisotropy Probe (WMAP) Observations:
  %Parameter Estimation Methodology,''
  Astrophys.\ J.\ Suppl.\  {\bf 148} (2003) 195
  [arXiv:astro-ph/0302218].
  %%CITATION = APJSA,148,195;%%

\bibitem{lucy74}
  L.~B.~Lucy,
  %``An Iterative Technique For The Rectification Of Observed Distributions,''
  Astron.\ J.\  {\bf 79} (1974) 745.
  %%CITATION = ANJOA,79,745;%%

\bibitem{rich72} 
  B.~H.~Richardson, 
  J. Opt. Soc. Am., {\bf 62}, 55 (1972).

\bibitem{baug_efs93} 
  C.~M.~Baugh and G.~Efstathiou, 
  Mon.Not.Roy.Astron.Soc. {\bf 265}, 145 (1993).

\bibitem{baug_efs94} 
  C. M. Baugh and G. Efstathiou,
  Mon.Not.Roy.Astron.Soc. {\bf 267}, 323 (1994).

\bibitem{Daubechies} 
  I. Daubechies, Ten Lectures on Wavelets, SIAM, Philadelphia, PA,
  1992. Vol.64, CBMS-NSF Conference Series in Applied Mathematices.

\bibitem{Wandelt:2000av}
  B.~D.~Wandelt, E.~Hivon and K.~M.~Gorski,
  %``The Pseudo-$C_l$ method: Cosmic microwave background anisotropy power
  %spectrum statistics for high precision cosmology,''
  Phys.\ Rev.\  D {\bf 64} (2001) 083003
  [arXiv:astro-ph/0008111].
  %%CITATION = PHRVA,D64,083003;%%

\end{thebibliography}
\end{document}